\documentclass[prb,showpacs,floatfix,twocolumn]{revtex4}
\usepackage{graphicx}
\usepackage{dcolumn}

\begin{document}

\newcommand*{\cm}{cm$^{-1}$\/}
\newcommand*{\TBCCO}{Tl$_2$Ba$_2$CaCu$_2$O$_{8+\delta}$\/}
\newcommand*{\YBCO}{YBa$_2$Cu$_3$O$_{7-\delta}$\/}
\newcommand*{\LSCO}{La$_{2-x}$Sr$_x$CuO$_{4+\delta}$\/}
\newcommand*{\BSCCO}{Bi$_2$Sr$_2$CaCu$_2$O$_{8+\delta}$\/}
\newcommand*{\BSCO}{Bi$_2$Sr$_2$CuO$_{6+\delta}$\/}
%
\title{Strong electron-Boson coupling effect in the infrared spectra of \TBCCO}
%
%

\author{N. L. Wang}
\email{nlwang@aphy.iphy.ac.cn}%
\author{P. Zheng}
\author{J. L. Luo}
\author{Z. J. Chen}
\affiliation{Key Laboratory of Extreme Conditions Physics,
Institute of Physics, Chinese Academy of Sciences, P.~O.~Box 603,
Beijing 100080, P.~R.~China}

\author{S. L. Yan}
\author{L. Fang}
\author{Y. C. Ma}
\affiliation{Department of Electronics, NanKai University, Tianjin
300071, P.~R.~China}
%
%
%
\begin{abstract}
We report on a study of the in-plane infrared response of a \TBCCO
film with $T_c$=108 K. The fingerprint evidence for a strong
electron-Boson coupling has been observed. The raw reflectance
below T$_c$ exhibits a knee structure at around 650 \cm and the
ratio of the reflectances below and above T$_c$ displays a
pronounced minimum at around 1000 \cm. In particular, the features
appear at higher energy scale than other bilayer cuprates with
T$_c$ around 90-95 K. The gap amplitude and the Boson mode energy
were extracted from the analysis of the bosonic spectral function.
In comparison with several other optimally doped high-Tc cuprates,
we find that the gap size scales with T$_c$, but Boson mode energy
has a tendency to decrease.
\end{abstract}


\maketitle

%
%
Superconductivity requires the electrons being glued together in
pairs. In conventional superconductors the pairing is due to the
coupling between electrons and phonons. Although the Boson
responsible for mediating the pairing or the "glue" in high
temperature superconducting cuprates (HTSC) remains elusive,
evidence for strong electron-Boson coupling effects began to
accumulate in various charge spectroscopies. In particular, many
of them are suggestive of a magnetic mode to glue the charge
carriers together. For example, the photoemission line shape near
($\pi$,0) in many of the HTSC in the superconducting state is
characterized by a sharp peak, followed at higher energy by a dip
and broad hump.\cite{Dessau} As one moves towards the zone
diagonal, a kink in the dispersion
develops.\cite{Kaminski,Lanzara} The kink appears at the same
energy scale as the dip.\cite{Kaminski} These anomalies were
widely believed to result from the interaction of electrons with a
collective mode which appears below T$_c$. The inferred mode
energy and its temperature dependence agree well with the magnetic
resonance observed at ($\pi,\pi$) by neutron
scattering.\cite{Ding,Norman,Campuzano} The coupling of the
electrons with magnetic mode has also been indicated in the
tunneling spectra of \BSCCO (Bi2212), where a sharp dip was
observed at a voltage beyond the gap edge
2$\Delta$.\cite{Zasadzinski}

Study of electron-phonon coupling effects in infrared spectra of
conventional superconductors began in early 1970's.\cite{Joyce}
Progresses have been made in recent years. Marsiglio \textit{et
al.}\cite{Marsiglio} showed that the electron-phonon spectral
function $\alpha^2F(\Omega)$ is closely related to the optical
scattering rate
\begin{equation}
   {1\over\tau(\omega)}=
   {\omega_p^2\over4\pi}Re{1\over\sigma(\omega)}
   ={\omega_p^2\over4\pi}{\sigma_1(\omega)\over{\sigma_1^2(\omega) +
   \sigma_2^2(\omega)}},
\label{chik}
\end{equation}
where $\omega_p$ is the plasma frequency. They found that
\begin{equation}
   W(\omega)=
   {1\over2\pi}{d^2\over d\omega^2}[\omega{1\over\tau(\omega)}]
\label{chik}
\end{equation}
is very close to $\alpha^2F(\Omega)$ in the phonon region with
additional wiggles beyond the phonon cutoff. However, no
correlation between the structure in $W(\omega)$ and phonon
density of states has been established for high-T$_c$ cuprates.
Instead, signature of the spin resonance mode in neutron
experiment is also indicated in the in-plane infrared spectrum.
Carbotte \textit{et al.}\cite{Carbotte} analyzed optical
conductivity $\sigma(\omega)$ in magnetically mediated d-wave
superconductors and argued that $W(\omega)$ of equ.(2) still gives
a good approximation to the charge-spin spectral density and that
the spin resonance feature should show up in W($\omega$) as a
result of coupling of the electrons to this magnetic mode. In
particular, the peak in W($\omega)$ below T$_c$ is correlated to
$\Delta+\Omega$, where $\Omega$ is the energy of the spin
resonance mode.\cite{Bourges} In another theoretical study of
optical conductivity in superconductors with quasiparticles
strongly coupled to their own collective spin modes, Abanov et
al.\cite{Abanov} emphasized that a deep minimum locating at
2$\Delta+\Omega_s$ in $W(\omega)$ is more relevant to
superconductivity. They also identified two weaker high frequency
singularities at 4$\Delta$ and 2$\Delta+2\Omega$. More recently,
Shulga \cite{Shulga} pointed out that the minimum in the ratio of
the raw reflectance $R_s(\omega)/R_n(\omega)$ is the key evidence
for the Boson-mediated superconductivity. He showed that the
minimum locates at frequency of 2$\Delta+\Omega$, where the
scattering rate displays a peak. Obviously, identifying these
features in the infrared spectra represents the fingerprint
evidence for the electron-Boson coupling in high-T$_c$ cuprates.

In this paper, we report on a study of the in-plane infrared
response of a \TBCCO (Tl2212) film with $T_c$=108 K. The above
mentioned features in reflectance spectra are clearly observed,
providing unambiguous evidence for the strong electron-Boson
coupling effect in this bilayer cuprate with higher T$_c$. In
particular, the features appear at higher energy scale than other
bilayer cuprates with T$_c$ around 90-95 K. We compared the
results on Tl2212 with other high-T$_c$ cuprates, and suggest that
a magnetic resonance mode exists in this bilayer material.

Epitaxial \TBCCO films were grown on LaAlO$_3$ substrates by dc
magnetron sputtering and a post-annealing process. X-ray
diffraction measurements confirm that the films were strongly
textured with the c-axis parallel to the c-axis of the substrate,
as shown in Fig. 1. Details of the film growth were described
elsewhere. \cite{Yan} Optimally doped films with thickness of 4000
$\AA$ were used for optical reflectance measurements. The
reflectance measurements from 50 to 25000~cm$^{-1}$ for
$\textbf{E}\parallel$ ab-plane were carried out on a Bruker 66v/S
spectrometer with the samples mounted on optically black cones in
a cold-finger flow cryostat using an \textit{in situ} overcoating
technique. \cite{Homes} The optical conductivity spectra were
derived from the Kramers-Kronig transformation.

%
%
\begin{figure}[t]
\centerline{\includegraphics[width=2.7in]{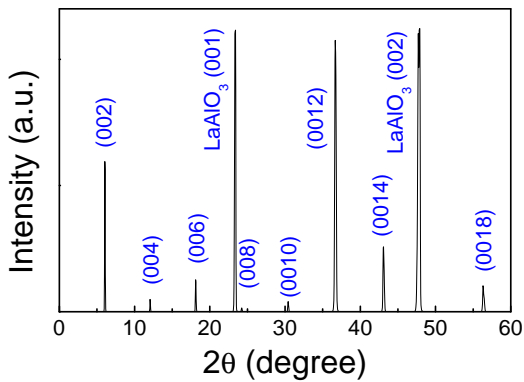}}%
\vspace*{-0.2cm}%
\caption{The X-ray diffraction pattern for a Tl2212 film.
All the sharp peaks can be assigned to the (00l) peaks of Tl2212 except those of LaAlO$_3$ substrate.}%
\label{fig1}
\end{figure}

Figure 2 shows the temperature dependent resistivity of the Tl2212
film. The zero-resistance temperature \textit{T}$_{c0}$ is 108 K,
which is much higher than other optimally doped bilayer cuprates
with \textit{T}$_{c0}\sim$ 90-95 K. The raw reflectances
$R(\omega$) from 50 to 2000 \cm and the calculated optical
conductivities $\sigma_1(\omega)$ at different temperatures are
displayed in Fig. 3(a) and 3(b), respectively. It should be
pointed out that the thickness of 4000 $\AA$ of our films appears
sufficient to avoid signals from the substrate. For this purpose,
we show, in the inset of Fig. 3(b), the far-infrared reflectance
spectrum measured on LaAlO$_3$ substrate at room temperature. The
spectrum is dominated by phonon peaks. However, no clear features
at the energies corresponding to the phonon peaks could be
detected in $R(\omega$) of the Tl2212 film.

%
%
\begin{figure}[t]
\centerline{\includegraphics[width=2.6in]{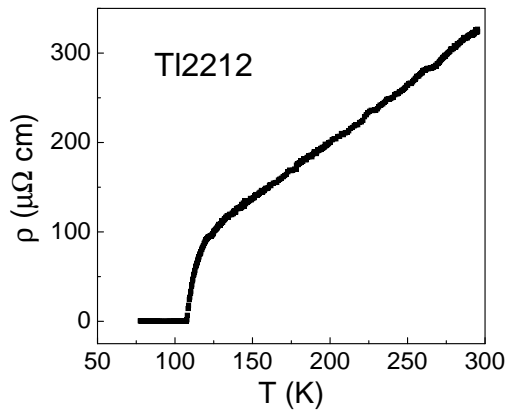}}%
\vspace*{-0.2cm}%
\caption{The resistivity vs. temperature curve for the film with zero resistivity at 108 K}%
\label{fig2}
\end{figure}
%

%
\begin{figure}[t]
\centerline{\includegraphics[width=2.7in]{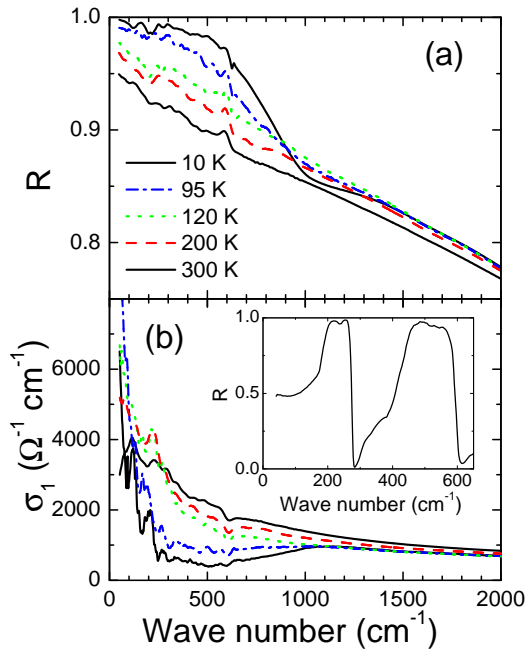}}%
\vspace*{-0.2cm}%
\caption{ab-plane optical data from 50 to 2000 \cm of the
optimally doped Tl-2212 with T$_c$=108 K. (a) the
temperature-dependent reflectance and (b) the
temperature-dependent $\sigma_1(\omega)$.
Inset: the far-infrared reflectance of the substrate LaAlO$_3$ at room temperature.}%
\label{fig3}
\end{figure}

In the normal state, the reflectance spectra show roughly linear
frequency dependence. With decreasing temperature, the
low-$\omega$ $R(\omega$) increases, being consistent with the
metallic behavior of the sample. At 10 K in the superconducting
state, $R(\omega$) shows a knee structure at around 650 \cm. Above
this frequency, the reflectance drops fast and becomes lower than
the normal-state values at around 1000 \cm. $R(\omega$) recovers
the linear-frequency dependence at higher frequencies. Similar but
weak behaviors were seen at 95 K, which is close to T$_c$. The
change of the $R(\omega)$ in the superconducting state with
respect to the normal state could be seen more clearly from the
plot of the ratio of the reflectance below T$_c$ over that above
T$_c$. Fig. 4(a) shows the ratio of
$R_{10K}(\omega)$/$R_{120K}(\omega)$ as a function of frequency.
It becomes evident that a maximum at around 650 \cm and a minimum
at around 1000 \cm exist in the plot. The scattering rate
1/$\tau(\omega)$ spectra at 10 K and 120 K, extracted from equ.
(1) with the use of plasma frequency of 1.6$\times 10^3$ \cm
determined by summarizing the optical conductivity up to 1 eV, are
displayed in Fig. 4(b). We can see that, at frequencies
corresponding to the maximum and minimum in the
$R_{10K}(\omega)$/$R_{120K}(\omega)$ plot, the 1/$\tau(\omega)$ at
10 K exhibits a rapid rise and a peak or a substantial overshoot
of the normal state result, respectively.

Obviously, the features caused by strong electron-Boson coupling
as discussed by Shulga \cite{Shulga} were observed in infrared
spectra of the optimally doped \TBCCO with T$_{c0}$=108 K. The
present work is very suggestive that a simple deviation or
suppression from the linear-frequency dependence in
1/$\tau(\omega)$, \textit{i.e.} the so-called "pseudogap" feature
in ab-plane optical response, may not be a direct indication for
the Boson mode being coupled to the electronic spectra. The dip in
the reflectance ratio $R_s/R_n$ or the "overshoot" in
1/$\tau(\omega)$ is more essential. We note that the overshoot
feature was observed previously on optimally doped \YBCO (YBCO) by
Basov et al. \cite{Basov1} and HgBa$_2$Ca$_2$Cu$_3$O$_{8+\delta}$
(Hg1223) by McGuire et al. \cite{McGuire}. The features in Bi2212
are weak, but a recent study by Tu et al. \cite{Tu} made it clear
that those features exist. However, the "overshoot" behavior is
absent in \LSCO (LSCO), although the suppression of low-$\omega$
scattering rate is still observed. \cite{Startseva}

%
\begin{figure}[t]
\centerline{\includegraphics[width=2.5in]{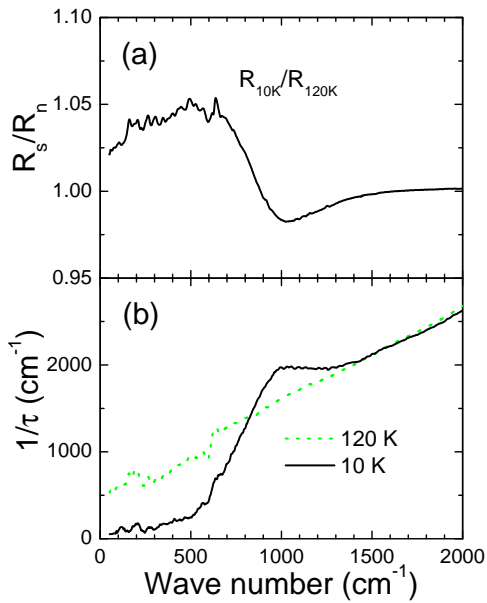}}%
\vspace*{-0.2cm}%
\caption{(a) The reflectance ratio below and above T$_c$ as a
function of frequency. (b) The frequency-dependent scattering rate
1/$\tau(\omega)$ spectra at 10 K and 120 K.}%
\label{fig4}
\end{figure}

Another remarkable observation here is that the "overshoot"
feature in 1/$\tau(\omega)$ appears at higher energy than the
corresponding features in YBCO and Bi2212, which are in the range
of about 800 - 900 \cm. \cite{Basov1,Tu} We note that the feature
in Hg1223 with T$_c$=130 K locates at even higher frequency beyond
1100 \cm. \cite{McGuire} Therefore, there is a tendency that the
energy of the feature scales with the superconducting transition
temperature.

%
\begin{figure}[t]
\centerline{\includegraphics[width=2.6in]{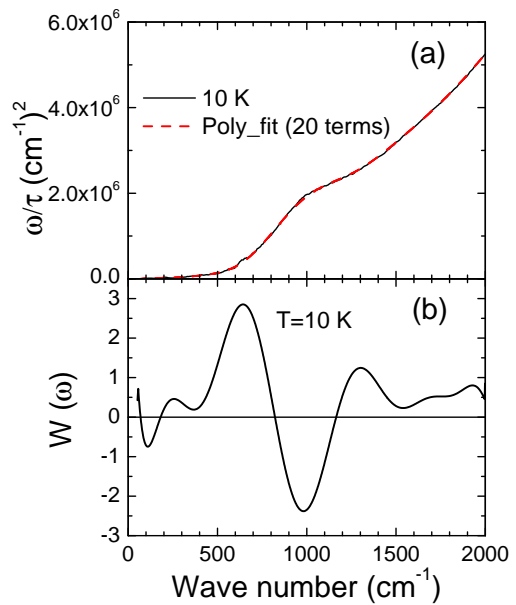}}%
\vspace*{-0.2cm}%
\caption{(a) The experimental data of $\omega/\tau(\omega)$ at 10
K together with a 20-term polynomial fit. (b) The bosonic spectral
function W($\omega$) vs. frequency derived from the polynomial fit curve.}%
\label{fig5}
\end{figure}

There are two different opinions about the origins of the Bosons
seen in charge spectroscopies. One is phonon origin \cite{Lanzara}
and the other the magnetic mode origin. We found that the phonon
scenario is inconsistent with the present experiment, for the
feature is seen only in the superconducting state, whereas phonons
exist also in the normal state. Recent study on the oxygen isotope
effect in the ab-plane optical properties of YBCO also ruled out
the phonons as the main player. \cite{Wang1} By contrast, Tl2212
shares the same coupling behavior as the YBCO and Bi2212, where
suggestions for the magnetic origin were proposed from the
inversion of the ab-plane optical conductivity.
\cite{Carbotte,Wang2,Tu} Note that, neutron experiments revealed a
main difference for the resonances in optimally doped YBCO and
Bi2212: the resonance peak in the later material exhibits a
considerable broadening in both energy and wave vector, which was
interpreted to be caused by the intrinsic inhomogeneities in
Bi2212. \cite{Burges2} Therefore, We can ascribe the weak "dip" or
"overshoot" features in Bi2212 to the broadening of the resonance
peak. The absence of those features in LSCO might be related to
the absence of the magnetic resonance. Since the "overshoot"
feature in Tl2212 is prominent, like the case in YBCO, we suggest
that a sharp magnetic resonance mode exists in this material.

The above discussion on the dip or "overshoot" behavior appears at
frequency of 2$\Delta+\Omega$, which contains the superconducing
gap and a Boson mode. It would be much desired to separate the two
different energy scales. This might be achieved from the analysis
of the bosonic spectral function W($\omega$) defined in equ.(2).
Since W($\omega$) is obtained by taking the second derivative of
the function $\omega/\tau(\omega)$, a smoothing of the
experimental data is required. Tu et al. \cite{Tu} suggested an
unambiguous method to extract W($\omega$) by fitting the
experimental measured quantity $\omega/\tau(\omega)$ with
high-order polynomial. This method has been used in present
analysis. In Fig. 5 (a), the $\omega/\tau(\omega)$ at 10 K is
shown together with a 20-term polynomial fit. The resulting
spectral function W($\omega$) vs. frequency is displayed in Fig.5
(b). We can see a large maximum at about 650 \cm, a deep negative
minimum at 1000 \cm, and a weak maximum at around 1300 \cm.
According to Carbotte et al., \cite{Carbotte} the large maximum
corresponds to $\Delta+\Omega$. In terms of Abanov et al.
\cite{Abanov}, the other two features correspond to
2$\Delta+\Omega$ and 2$\Delta+2\Omega$, respectively. Then, we
obtain $\Delta$=350 \cm (or 43 meV) and $\Omega$=300 \cm (or 37
emV) for \TBCCO.

We find obviously that the extracted gap $\Delta$ is larger than
the values for Bi2212 (35 meV) and YBCO (28 meV), but the Boson
mode energy $\Omega$ is smaller than the corresponding values of
43 meV for Bi2212 and 41 meV for YBCO. \cite{Tu,Carbotte} Since
the T$_c$ of Tl2212 is very close to that of Bi2223, it would be
interesting to compare the result with Bi2223. The gap amplitudes
for n=1, 2, 3 compounds in Bi-based family have been studied in
photoemission experiments by Feng et al. \cite{Feng} They found
that the leading-edge-midpoint (peak position) gap values are 10
(21), 24 (40), 30 (45) meV for n=1, 2, 3 systems, respectively. It
is seen that the gap values for Bi2212 and Tl2212 extracted from
in-plane optical data are more closer (but a bit smaller) to their
peak position gap values for n=2 and 3 systems. In Fig.6, we plot
$\Delta$, $\Omega$ and the energy of "overshoot"
(2$\Delta+\Omega$) versus T$_c$ for several optimally doped
cuprates where those data are available. We find that the gap
increases with T$_c$, but the Boson mode energy has the tendency
to decrease with T$_c$. In the figure, the data of YBCO appear to
deviate substantially from the scaling behavior. The reason is not
clear. One possibility is that the existence of the Cu-O chains,
which also become superconducting below T$_c$, helps to enhance
T$_c$, even though its gap value is smaller than Bi2212 and
Tl2201.

%
\begin{figure}[t]
\centerline{\includegraphics[width=2.6in]{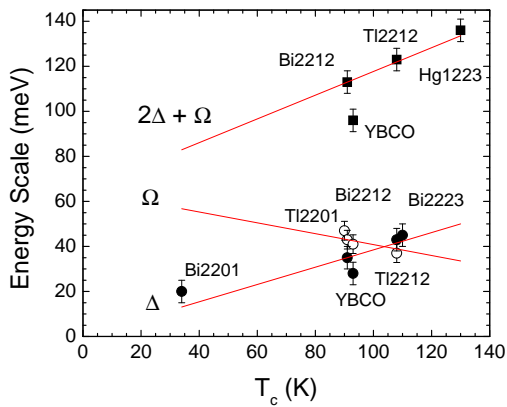}}%
\vspace*{-0.2cm}%
\caption{Plots of $\Delta$, $\Omega$ and the energy of overshoot
(2$\Delta+\Omega$) versus T$_c$ for several optimally doped
cuprates. The data for Bi2212\cite{Tu} and YBCO\cite{Carbotte} are
from inversion of optical spectra. The $\Delta$ values for Bi2201
and Bi2223 are from photoemission study (peak position
gap).\cite{Feng} Hg1223 data is from \cite{McGuire}. The $\Omega$
value for Tl2201 is from neutron experiment.\cite{He} The straight
lines are for eye guidance. We start
to put straight lines for (2$\Delta+\Omega$) and $\Omega$, the third one for $\Delta$ is then extracted.}%
\label{fig6}
\end{figure}

Earlier studies have established that the T$_c$ in optimally doped
cuprates is proportional to the condensed carrier density n$_s$.
The present study shows that T$_c$ is also correlated with the
energy scale of the superconducting gap $\Delta$. Within current
understanding, both the n$_s$ and $\Delta$ are important
quantities characterizing superconducting state. $\Delta$ reflects
the pairing strength, while n$_s$ is an indication of the phase
stiffness of the pairing.\cite{Feng,emery} To achieve maximum
T$_c$, both the superconducting gap and the condensate should be
maximized.

Finally, we briefly comment on the two sizeable features at 600
\cm and 230 \cm, which are present in all measured temperatures.
Since the frequencies are away from the phonon modes of the
substrate,\cite{note} they are not likely to come from the
substrate due to the possible leakage of the film in the long wave
length region. The high-frequency mode was seen in other high-Tc
cuprates, e.g. Bi2212\cite{Tu} and YBCO\cite{Basov1,Wang1}, and
was assigned to the transverse optic phonon (Cu-O stretching
mode). The 230 \cm mode was also reported in earlier reflectance
study on YBCO, where it was suggested to be of electronic origin
for the spectral weight of the mode was found to be an order of
magnitude larger than what is expected for a
phonon.\cite{Bernhard}. Comparing the present data with those
studies, we found that the high-frequency mode appears as an
antiresonance rather than a peak in the conductivity spectra, and
the strength of the 230 \cm mode is not weakened very much at high
temperature, which is in contrast to the reported behavior in
YBCO. Further studies are required to illustrate the difference.

In summary, we have presented a set of high-quality in-plane
optical data for \TBCCO (Tl2212) with $T_c$=108 K. The fingerprint
evidence for strong electron-Boson coupling effect has been
observed in reflectance spectra for this bilayer cuprate with
higher T$_c$. In particular, the energy scale of the feature
appears at higher energy than other bilayer cuprates with T$_c$
around 90-95 K. The gap amplitude and the Boson mode energy were
extracted from the analysis of the bosonic spectral function. It
is found that the gap size scales with T$_c$ for different
cuprates. We discussed the origin of the Boson and compared the
results on Tl2212 with other high-T$_c$ cuprates.

This work is supported by National Science Foundation of China
(No.10025418, 19974049), the Knowledge Innovation Project of
Chinese Academy of Sciences.
%
%

\end{document}